\documentclass[a4paper]{PoS}

\usepackage{amsmath}
\usepackage{amssymb}
\usepackage{graphicx}
\usepackage{amsfonts}
\usepackage{bbm}
\usepackage{color}

\usepackage[normalem]{ulem}
\usepackage{slashed}
\renewcommand{\imath}{\text{i}}

\newcommand{\T}{\textsf{T}}
\newcommand{\diag}[1]{\text{diag}(#1)}

\newcommand{\Eq}[1]{Eq.~(\ref{#1})}
\newcommand{\Fig}[1]{Fig.~\ref{#1}}

\newcommand{\M}{M}

\title{Color-superconductivity and \\inhomogeneous chiral symmetry breaking in isospin-asymmetric quark matter}

\ShortTitle{CSC and inhomogeneous $\chi$SB at nonzero $\mu_I$}

\author{\speaker{Daniel Nowakowski}\\
        Theoriezentrum, Institut f\"ur Kernphysik, TU Darmstadt, Schlossgartenstra\ss{}e 2, 64289 Darmstadt, Germany\\
        E-mail: \email{danielno@theorie.ikp.physik.tu-darmstadt.de}}

\author{\speaker{Stefano Carignano}       \\
INFN, Laboratori Nazionali del Gran Sasso, Via G. Acitelli, 22, 67100 Assergi (AQ), Italy
        E-mail: \email{carignano@lngs.infn.it}}

\abstract{
We investigate the effects of isospin asymmetry on the competition between color-superconductivity and inhomogeneous chiral symmetry breaking in dense two-flavor quark matter
using an extended Nambu--Jona-Lasinio model. We confirm the appearance of a coexistence window where chiral symmetry is inhomogeneously broken and a nonzero spatially homogeneous diquark gap is present, consistently with previous works, and show that such a phase survives at nonzero isospin chemical potentials. We also discuss how the model phase structure becomes modified as large isospin asymmetries are considered. 
    }

\FullConference{The Modern Physics of Compact Stars 2015\\
		30 September 2015 - 3 October 2015\\
		Yerevan, Armenia}

\begin{document}

\section{Introduction}

The study of the properties of strong interactions under extreme conditions is one of the most challenging tasks in contemporary nuclear physics. While heavy-ion experiments and ab-initio lattice simulations of quantum chromodynamics (QCD) allowed to shed some light on the behavior of hadronic matter at high temperatures and vanishing baryon densities, the other side of the QCD phase diagram, characterized by low temperatures and high densities, is still poorly understood. 
While it is expected that at asymptotically high densities quark matter is deconfined and forms a homogeneous color-superconductor \cite{Barrois:1977xd,Bailin:1983bm} (the most favored pairing pattern being the so-called ``color-flavor locked'' one \cite{Alford:1998mk}), it is still not clear which phase structure is realized at the intermediate range where the density of the system reaches a few times nuclear matter density. 
Aside from peculiar states of nuclear matter, other possible candidates for the ground state of cold and dense hadrons could be quarkyonic matter \cite{McLerran:2007qj,Kojo:2009ha}, 
 color-superconducting states (for a dedicated review, see e.\;g. \cite{Alford:2007xm}), or phases where chiral symmetry is broken through the formation of crystalline quark-antiquark condensates. This latter idea, which has its precursors in the studies of pion condensation in nuclear matter, has recently received a significant amount of attention through a series of model studies, which seem to agree on the existence of an inhomogeneous phase at intermediate densities and low temperatures (see \cite{Broniowski:2011ef} for an historical overview on the idea of inhomogeneous chiral condensates, \cite{Buballa:2014tba} for a review on recent results and \cite{Buballa:2015awa} for 
 a discussion on some consequences of their existence within compact stars). 

While waiting for the next generation of heavy-ion experiments to probe this region, compact stars provide the only known laboratory in nature where such extreme densities could be reached.  If exotic phases are realized in their cores, they could dramatically alter the thermodynamical properties of these objects, possibly leading to strong experimental signatures. Astrophysical data from these sources could then be compared with predictions coming from different phenomenological models with the aim of determining the properties of cold and dense matter.
It is well known that in compact stars global electric neutrality must be satisfied. This condition, together with $\beta$-equilibrium, leads to an imbalance in the chemical potential for the  quark flavors due to their different charges. While a complete calculation of compact star matter would require the introduction of electrons and a self-consistent determination of the amount of imbalance through the neutrality condition, in this work we will focus on the effects of this asymmetry on quark matter by introducing an isospin chemical potential 
\begin{align}
\label{eq:muI}
\mu_I = \mu_u - \mu_d \,,
\end{align}
and by keeping it as a free external parameter.  The effects of isospin imbalance on inhomogeneous chiral symmetry breaking are non-trivial, and we defer to \cite{Nowakowski:2015ksa} and \cite{dno} for a detailed discussion. Of particular interest for us here instead is the study of the competition of color-superconductivity with inhomogeneous chiral symmetry breaking, a long-standing problem which has already been considered for isospin-symmetric matter. The original expectation presented in \cite{Park:1999bz,Shuster:1999tn} that color-superconductivity usually wins over inhomogeneous chiral symmetry breaking was already challenged in \cite{Rapp:2000zd}, and more recently it was found in \cite{Sadzikowski:2002iy, Sadzikowski:2006jq} that the two could coexist in a finite region of the phase diagram. 
All these studies nevertheless neglected the effects of isospin imbalance in the system, a gap which we aim to fill with the present work. 

Studies of color-superconductivity in isospin-asymmetric systems have been performed in several works (see e.\;g.\;\cite{Alford:2007xm} for a review). Limiting to the two light quark flavors, an imbalance on the quark chemical potentials shifts their respective Fermi surfaces apart, effectively disfavoring the BCS pairing mechanism, which involves quarks with equal momenta.
For sufficiently large asymmetries, the energy gain through the formation of the diquark gap is then not enough to compensate the cost of pulling quarks out of their respective Fermi surfaces, and beyond a critical stress chemical potential BCS pairing becomes disfavored \cite{Chandrasekhar,Clogston:1962zz}.  In this regime new types of color-superconducting phases might appear, including inhomogeneous ones (for reviews on crystalline color-superconductors see \cite{Casalbuoni:2003wh,Anglani:2013gfu}).

While of course the coexistence of inhomogeneous chiral symmetry breaking and inhomogeneous color-superconducting solutions in isospin asymmetric matter is a fascinating possibility, a study of such configurations would be extremely involved. For this, in the following we will limit ourselves to considering BCS-type spatially homogeneous diquark gaps.

\section{Model}
We consider deconfined quark matter within a two-flavor Nambu--Jona-Lasinio (NJL) model. In order to describe the competition of inhomogeneous chiral symmetry breaking with color-superconductivity in asymmetric matter we consider the Lagrangian density proposed in \cite{Asakawa:1989bq} and extend it by adding a diquark pairing channel. Our model is then given by 
\begin{align}
\mathcal{L}= \mathcal{L}_0+\mathcal{L}_A+\mathcal{L}_B+\mathcal{L}_C\label{eq:lagrangen} \,.
\end{align}
The first is a free Dirac part, 
\begin{align*}
  \mathcal{L}_0=\bar{\psi}(i \gamma^\mu\partial_\mu-\hat{m})\psi \,,
\end{align*}
where the quark fields $\psi$ are represented by a $4N_f N_c$ spinor and $\hat{m}=\diag{m_u,m_d}$ is the current mass matrix. 
The two interaction terms related to chiral symmetry breaking are
\begin{align}
 \mathcal{L}_A&=G_1\left(\left(\bar{\psi}\psi\right)^2+\left(\bar{\psi}\vec{\tau}\psi\right)^2+\left(\bar{\psi}\imath\gamma_5\psi\right)^2+\left(\bar{\psi}\imath\gamma_5\vec{\tau}\psi\right)^2\right) \,,\label{eq:lagrangeInter1} \\
 \mathcal{L}_B&=G_2\left(\left(\bar{\psi}\psi\right)^2-\left(\bar{\psi}\vec{\tau}\psi\right)^2-\left(\bar{\psi}\imath\gamma_5\psi\right)^2+\left(\bar{\psi}\imath\gamma_5\vec{\tau}\psi\right)^2\right)  \,, \label{eq:lagrangeInter2}
 \end{align}
 with $G_1,G_2$ dimensionful coupling constants and $\tau^a$ Pauli matrices in flavor space with the index $a$ running from 1 to 3. 
Considering only flavor-diagonal structures, we note that in $\mathcal{L}_A$ the quark flavors are completely decoupled, while the instanton-induced
 term $\mathcal{L}_B$ has the structure of a 't-Hooft determinant and effectively mixes the two. 
 For later convenience, following \cite{Frank:2003ve} we parameterize the two coupling $G_1$ and $G_2$ in terms of a common $G$ and a parameter $\alpha$ as
\begin{align}
  G_1=(1-\alpha)\;G,\quad G_2=\alpha\;G \,,
\end{align}
so that $\alpha =0$ describes the limit of decoupled quarks, while for $\alpha=0.5$ we have $G_1 = G_2$ and the maximal coupling is realized. 

Finally, the diquark pairing is described by the term
 \begin{align}
\mathcal{L}_C&=H\sum_{A=2,5,7} \left(\bar{\psi} i \gamma_5 C \tau_2 \lambda_A \bar\psi^\T\right)\left(\psi^\T C i \gamma_5 \tau_2\lambda_A \psi\right) \,,\label{eq:lagrangeInter3}
\end{align}
where $H$ is another dimensionful coupling constant,  $\lambda_A$ are  Gell-Mann matrices in color space (with color index $A$)
 and $C=i\gamma^2\gamma^0$  denotes the charge conjugation operator.

In order to determine the thermodynamic potential of the system, we perform the common mean-field approximation 
by expanding around the expectation values of the relevant condensates which will then become the variational parameters to be determined 
in obtaining the ground state of the system. 
In particular, we define the scalar and pseudoscalar chiral condensates
\begin{align}
       S_u(\vec x) & =\langle\bar{u}u\rangle  \,, \qquad S_d(\vec x)  =\langle\bar{d}d\rangle    \qquad \text{and} \\
       P_u(\vec x) & =\langle\bar{u}i\gamma_5u\rangle \,, \quad P_d(\vec x) =\langle\bar{d}i\gamma_5 d\rangle   \,.
\end{align}
In the following, we neglect condensates which are off-diagonal in flavor space. This in turn implies that we do not consider the possibility of charged pion condensation,
which is expected to set in as the isospin imbalance becomes sufficiently large to excite $\pi^\pm$ in the system. In order to be consistent with this approximation, in our numerical calculations we will limit ourselves to moderate values of $\mu_I$.

The study of inhomogeneous chiral symmetry breaking in isospin-asymmetric matter is rather involved, and some particular care is required when introducing an ansatz for the spatial dependence of the chiral condensate (for a detailed discussion, we refer the reader to \cite{dno}).
 In the following we will consider the simplest possible choice, namely we introduce spatially-dependent mass functions per quark flavor 
\begin{align}
  \label{eq:malpha}
  \hat{M}_f(\vec x)  = m_f-4 G \Big\{\big[S_f + i\gamma_5 P_f\big]  -\alpha\big[ (S_f - S_h) + i\gamma_5 (P_f + P_h) \big] \,\Big\} \,,
\end{align}
where $f,g \in \lbrace u, d \rbrace \,,  f \neq g $, and make a plane-wave ansatz for each flavor mass function 
\begin{align}
\hat M_f(\vec x)=\M_f \exp{\left(\imath \gamma^5 \vec q_f\cdot \vec x\right)},
\label{eq:Mofx}
\end{align}
where $\M_f$ denotes the amplitude and $\vec q_f$ is the wave vector of the modulation. We assume that $\vec q_u$ and $\vec q_d$ have the same magnitude, but opposite sign\footnote{This is necessary to obtain the correct CDW limit in isospin-symmetric matter \cite{dno}. }.

For the quark-quark interaction \Eq{eq:lagrangeInter3}  we focus on scalar condensates of the form 
\begin{align}
 s_{2A}=\langle \psi^\T C \gamma_5 \tau_2 \lambda_{A} \psi\rangle.
\end{align}
Since for $N_f=2$ the diquark condensates $s_{2A}$ (with $A=2,5,7$) form a vector in color-space, that can always be rotated by global $\text{SU}(3)$-color transformations, we focus on the $s_{22}$ direction and introduce the 2SC condensate
\begin{align}
 s_{22}=\langle \psi^\T C\gamma_5 \tau_2 \lambda_2 \psi\rangle
\end{align}
which, as in standard conventions, is made of red and green up and down quarks, while blue quarks remain unpaired.
 For later convenience we also define the gap associated with this 2SC condensate as
 \begin{align}
 \Delta=-2H\,s_{22} \,,
 \label{eq:deltagap}
\end{align}
and as in \cite{Sadzikowski:2002iy} we assume it to be spatially constant.

In presence of an isospin imbalance, the up and down quark chemical potentials will split. 
We then introduce the chemical potential in our equations as a diagonal matrix in flavor space
\begin{align}
 \hat{\mu}=\textrm{diag}\lbrace\mu_u,\mu_d\rbrace =  \textrm{diag}\left\lbrace \bar\mu+\frac{\mu_I}{2},\, \bar\mu-\frac{\mu_I}{2} \right\rbrace \,,
\end{align}
where we introduced the flavor-averaged chemical potential
\begin{align}
 \bar\mu=\frac{\mu_u+\mu_d}{2} \,.
\end{align}

As customary when dealing with diquark condensates, we employ the Nambu-Gor'kov (NG) formalism. Assuming that $\psi$ and $\psi^C=C\bar\psi^T$ are formally independent fields, we introduce the bispinors 
\begin{align}
 \Psi=\frac{1}{\sqrt{2}}\begin{pmatrix}\psi\\\psi^C\end{pmatrix}
\end{align}
and rewrite the mean-field model Lagrangian as  
\begin{align}
 \mathcal{L}_{MF}+\bar\psi \gamma_0 \hat\mu\psi =\bar\Psi S^{-1}\Psi-\mathcal{V}  \,,
\end{align}
where $\mathcal{V}$ is a field-independent term and $S^{-1}$ is the inverse quark propagator, given by
\begin{align}
  S^{-1}=\begin{pmatrix} \imath \slashed\partial - \hat M(\vec x) + \hat \mu \gamma^0 & \Delta \gamma_5 \tau_2\lambda_2\\ -\Delta^* \gamma_5\tau_2 \lambda_2 & -\imath\slashed\partial - \hat M(\vec x)-\hat\mu\gamma^0\end{pmatrix} \,.
\end{align}
Here we note that due to the isospin imbalance the diagonal elements of the propagator in NG space have a non-trivial flavor structure. This, in addition to the explicit space-dependence of the chiral condensate, significantly complicates our calculations.

At this point we are ready to calculate the thermodynamic potential per unit volume, which can be written using standard techniques \cite{Kapusta:2006pm} as
\begin{align}
\label{eq:Omega2}
\Omega(T,\lbrace\mu_f\rbrace;\, \lbrace{M_f\rbrace}, q, \Delta)
&= \Omega_{\rm kin} \, +\Omega_{\rm cond}\,,
\end{align}
where 
\begin{align}
\Omega_\text{kin}=- \frac{1}{2} T 
\sum_{n}
\mathrm{Tr}_{D,c,f,V} \, \mathrm{Log}\bigg(\frac{1}{T}S^{-1}(i\omega_{n},\vec p)\bigg) \,,
\end{align}
with fermionic Matsubara frequencies $\omega_n$. The factor $1/2$ is present due to the artificial doubling of degree of freedoms by using the Nambu-Gor'kov formalism. The inverse propagator in momentum space is given by
\begin{align}
 S^{-1}_{p_m,p_n}=\gamma^0\left(i\omega_{p_m}-\mathcal{H}_{p_m,p_n}\right)\delta_{\omega_{p_n},\omega_{p_n}}
\end{align}
in terms of the Matsubara frequencies and an effective Hamiltonian $\mathcal{H}$. For static condensates the propagator is diagonal in the frequencies and (assuming a sensible regularization of the functional trace) the Matsubara sum can be performed analytically, leading to
\begin{align}
\label{thermpot1}
  \Omega_\text{kin}=
  -\sum_{\lambda} \left[ |\lambda|+2T\,\mathrm{Log}\left(1+\exp{\left(-\frac{|\lambda|}{T}\right)}\right)\right] \,,
\end{align}
where the remainder of the trace has been expressed as a sum over the eigenvalues $\lbrace \lambda \rbrace $ of the quark Hamiltonian 
in NG, Dirac, color, flavor and momentum space.

The diagonalization of the Hamiltonian turns out to be rather tedious, but can be performed analytically thanks to the fact that blue quarks do not participate in the 2SC pairing.
After introducing energy projectors which allow us to separate the contributions from particles and antiparticles, 
we obtain for the red and green quarks  (to get the corresponding antiquark energies it is sufficient to invert the signs of both $\bar\mu$ and $\mu_I$)
\begin{align}
\label{eigenvalsRG}
\lambda^f_{r,g} =\sqrt{(\epsilon^f_\pm - \bar\mu)^2+\left|\Delta\right|^2} - s_f \frac{\mu_I}{2}  \,,
\end{align}
while the blue quark energies are simply 
\begin{align}
\label{eigenvalsB}
\lambda^f_b = \epsilon^f_\pm - \left( \bar\mu + s_f \frac{\mu_I}{2} \right) \,,
\end{align}
with $s_u = 1$, $s_d = -1$. For the case of our simple plane wave modulation, the diagonalization in momentum space can be performed through a chiral rotation (see e.g. \cite{Buballa:2014tba}) and the eigenvalue part related to inhomogeneous chiral symmetry breaking is given by 
\begin{align}
\label{eigenvals1}
\epsilon^f_\pm=\sqrt{\vec p^2+\M_f^2+\frac{ \vec q_f^2}{4}\pm\sqrt{\M_f^2\, \vec q_f ^2+\left(\vec q_f \cdot \vec p\right)^2}}  \;.
\end{align}
Our result naturally reduces to the one obtained in \cite{Sadzikowski:2002iy} in the limit $\mu_I \to 0$. 
 In order to remove the divergencies in \Eq{thermpot1}, we employ a Pauli-Villars inspired scheme where we replace only the $\epsilon$ in the temperature-independent part of the thermodynamic potential by their regularized counterparts,
\begin{align}
 \epsilon_\pm^f\rightarrow \sum_{j=0}^3 c_j \sqrt{\left(\epsilon_\pm^f\right)^2+j \Lambda^2} \,,
\end{align}
where $c_0=-c_3=1, c_2=-c_1=3$ and $\Lambda$ is our cutoff.  This peculiar prescription, which reduces to the one introduced in \cite{Nickel:2009wj} for $\Delta = \mu_I = 0$,
 has the advantage of not introducing unphysical medium dependencies of 
the variational parameters which would violate the Silver-Blaze property.

The condensate part absorbs all field-independent terms and reads 
\begin{align}
 \Omega_\text{cond}= \frac{1}{V}\int_V d^3x \, \mathcal{V} =
\frac{1}{8G}\;\frac{1}{1-2\alpha}\left[ \left(1-\alpha\right) \left(\M_u^2+\M_d^2\right)-2\alpha \M_u\M_d\right]
+\frac{|\Delta|^2}{4H} \,,
\end{align}
where we used the fact that $\vec q_u=-\vec q_d$.

After having derived an expression for the mean-field thermodynamic potential, we are now ready to determine 
 the variational parameters for our ansatz of spatially modulated quark masses and an homogeneous diquark gap, namely $(\M_u,\M_d,q,\Delta)$ (cfr. Eqs.\;\ref{eq:Mofx} and \ref{eq:deltagap}).

\section{Numerical results}

In this section we present numerical results of our study. For our calculations we employed $\alpha = 0.2$, 
which should provide
a realistic degree of flavor mixing \cite{Frank:2003ve}, together with $\Lambda=728.4\,\text{MeV}$ and $G\Lambda^2=6.6$ in our Pauli-Villars regularization scheme.  In addition, we will work in the chiral limit by setting the current quark masses $m_u$ and $m_d$ to zero\footnote{While strictly speaking this would imply that the pion mass is zero, here we use the chiral limit simply as a useful approximation for evaluating the thermodynamic potential, and assume that no pion condensation occurs for $\mu_I$ below the physical pion mass of $m_\pi \sim 135$ MeV.}.  This gives a constituent quark mass in vacuum of $M_{vac} = 330 $ MeV.   The coupling $H$ will be treated as a free parameter and typically  fixed to $G/2$.
Without loss of generality, we will consider only positive isospin chemical potentials, $\mu_I \geq 0$.  All our calculations have been performed at zero temperature.

\subsection{Isospin-symmetric matter}

\begin{figure}
  \centering
  \includegraphics[width=0.48\textwidth]{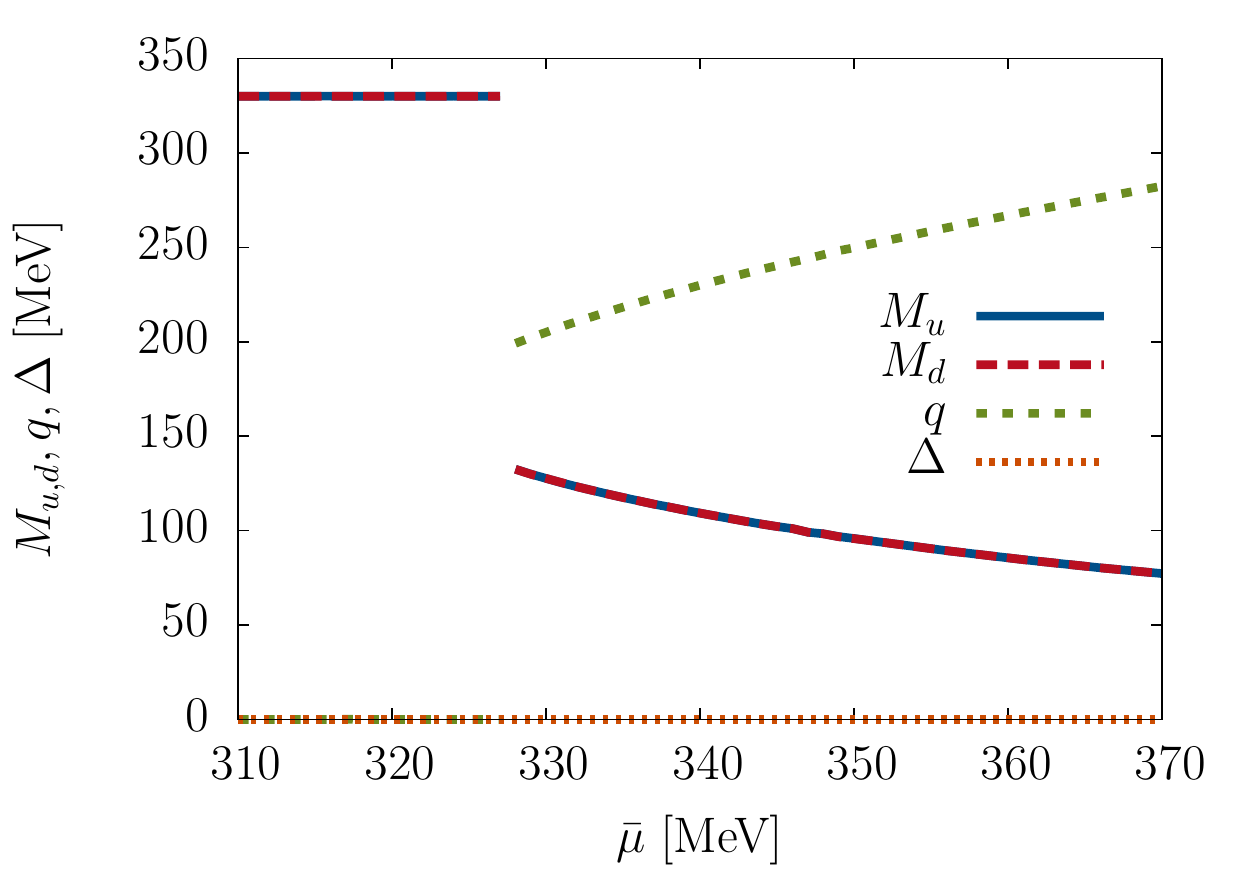}
  \includegraphics[width=0.48\textwidth]{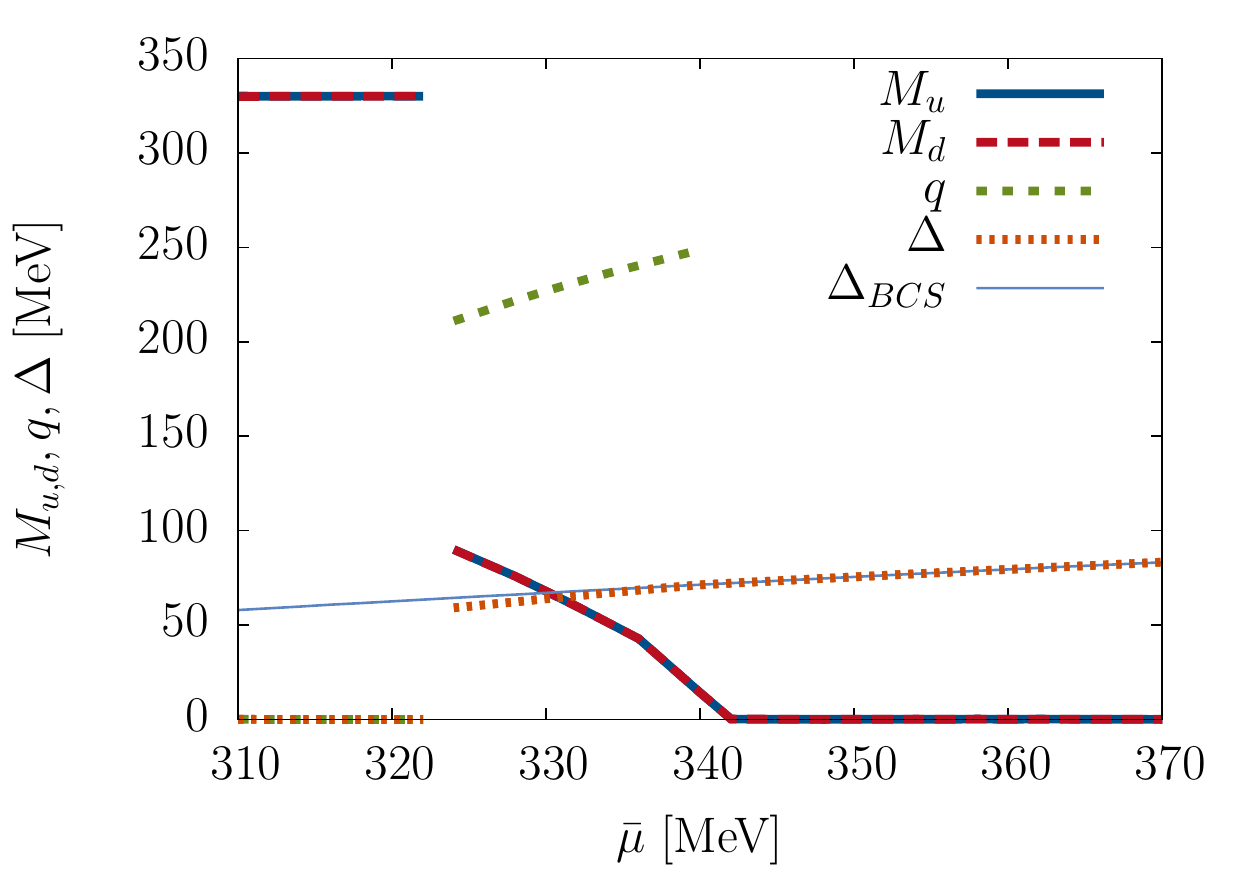}
  \caption{Energetically favored values of the amplitudes $\{M_f\}$, the wave-number $q$ for
    inhomogeneous chiral condensates and the diquark gap $\Delta$ for $T = \mu_I = 0$. 
     Left panel: $H=0$. Right: $H= G/2$.
         For comparison, in the right panel the value of the diquark gap for a pure 2SC solution in absence of chiral symmetry breaking is also shown.
      \label{fig:mui0one}}
\end{figure}

\begin{figure}
   \centering
   \includegraphics[width=0.48\textwidth]{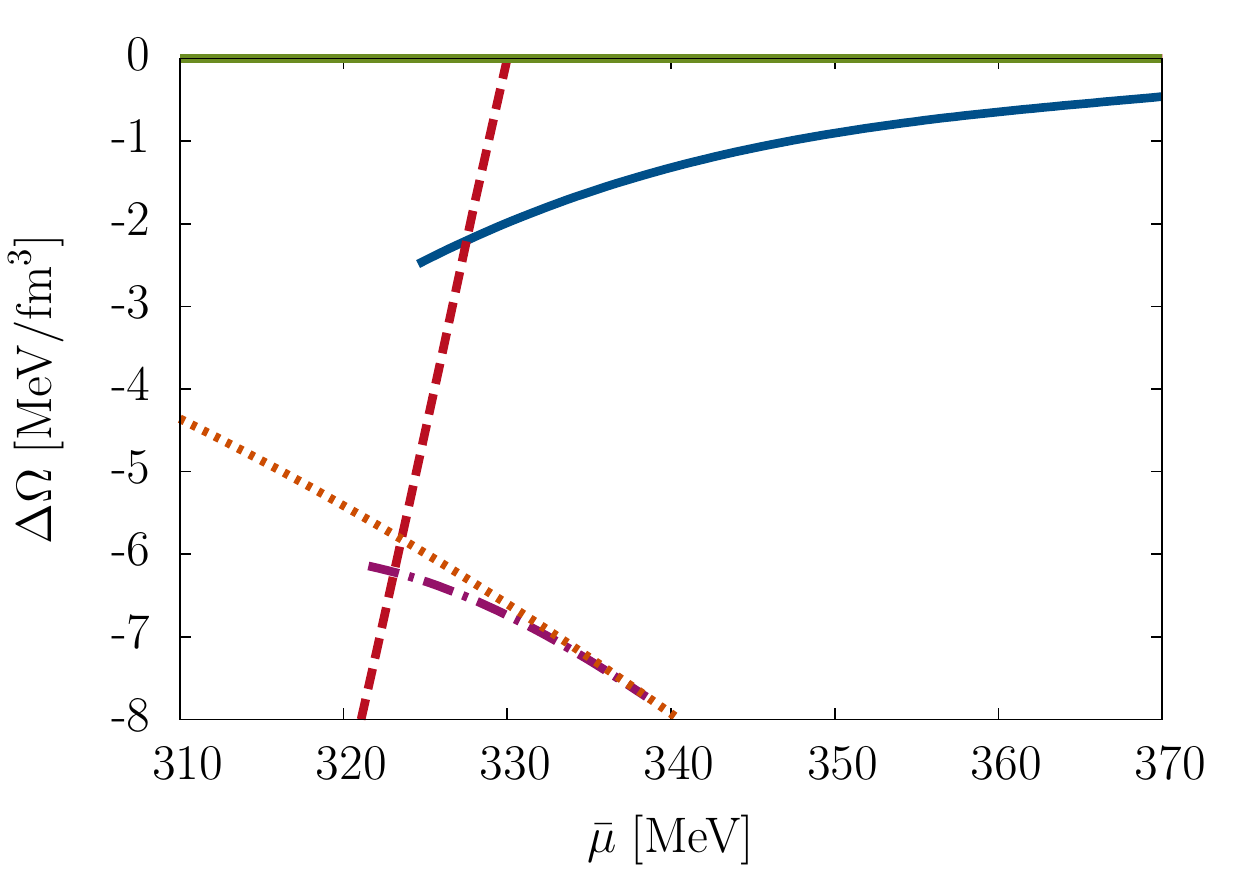}
   \caption{Thermodynamic potential (normalized to the restored phase) as a function of the average quark chemical potential at $T= \mu_I=0$ and $H=G/2$ for homogeneous chiral condensates (dashed red), inhomogeneous chiral condensates (solid blue), only diquark condensation (dotted orange) and the coexistence phase with simultaneous inhomogeneous chiral symmetry breaking and homogeneous color-superconductivity (dash-dotted purple). \label{fig:deltaomegamui0}}
\end{figure}

Before starting our discussion on isospin-asymmetric matter, in order to set the stage we first study the interplay of inhomogeneous chiral symmetry breaking with homogeneous color-superconductivity in the $\mu_I = 0$ case. 
As a first step, in the left panel of \Fig{fig:mui0one} we show the order parameters related to the chiral condensate as a function of the average quark chemical potential $\bar\mu$ for vanishing $H$. In this case there is no diquark gap and we only observe a first-order transition (as expected when dealing with a plane wave modulation) at $\mu \lesssim M_{vac}$ from the vacuum to a phase where chiral symmetry is inhomogeneously broken. This crystalline phase extends up to arbitrarily high chemical potentials, giving rise to the inhomogeneous ``continent'' discussed in \cite{Carignano:2011gr}.   If we now turn on the diquark coupling, we observe that the system undergoes a transition from the vacuum into a coexistence phase where both a crystalline chiral condensate and a nonzero diquark gap form, as can be seen from the right panel in \Fig{fig:mui0one}, where results for $H= G/2$ are shown. The onset of this phase, which is again first-order,
occurs at a slightly lower chemical potential ($\bar\mu \approx 325$ MeV) compared to the $H=0$ case: this is easily understood, since the formation of a diquark gap leads to an additional energy gain against the chirally broken homogeneous phase, as can be seen from \Fig{fig:deltaomegamui0}, where the thermodynamic potentials associated with the different phases (and normalized with respect to $\Omega(M_u = M_d = \Delta = 0)$ of the restored phase) are shown. 
The biggest difference in this case however is that the inhomogeneous phase does not extend to arbitrarily high chemical potentials: instead, the amplitude of the chiral condensate decreases to zero, so that the system reaches a chirally restored phase with a nonzero diquark gap (what we will refer to in the following as a "pure 2SC phase"). At this second-order transition, which occurs at $\bar\mu \approx 342 $ MeV,  the amplitude of the chiral condensate smoothly reaches zero and the diquark gap does not jump.

The size of the inhomogeneous window shrinks as the diquark coupling $H$ is increased. In particular, we observe that for our particular choice of model parameters and within our simple CDW-type ansatz for the chiral condensate, already at $H= 3/4 \, G$ the pure 2SC phase has become thermodynamically favored for all $\bar\mu$ over the coexistence one.

In the right panel of \Fig{fig:mui0one} we also show for comparison the value of the diquark gap $\Delta_{{BCS}}$ calculated for a pure 2SC solution in absence of chiral symmetry breaking. There one can see how the presence of a nonzero chiral condensate and consequently a lower density of the system weaken the gap, which in our case smoothly approaches the pure 2SC value as the amplitudes $M_f$ gradually melt to zero.

At this point we can briefly comment on the difference between our results and those presented in \cite{Sadzikowski:2002iy}.  There, the transition to the homogeneous pure 2SC phase occurs through a first-order transition as the chiral condensate drops sharply to zero.  We interpret this discrepancy as a result of the different regularizations employed in our two works, an issue which is likely unrelated to the effects of color-superconductivity on the behavior of the chiral order parameters. As a further evidence sustaining this statement, we observe that in \cite{Sadzikowski:2002iy,Sadzikowski:2006jq} the chiral restoration transition is first-order even in absence of a diquark gap, 
contradicting several recent results (see e.g. \cite{Buballa:2014tba}).

\subsection{Asymmetric matter}

\begin{figure}
 \centering
 \includegraphics[width=0.48\textwidth]{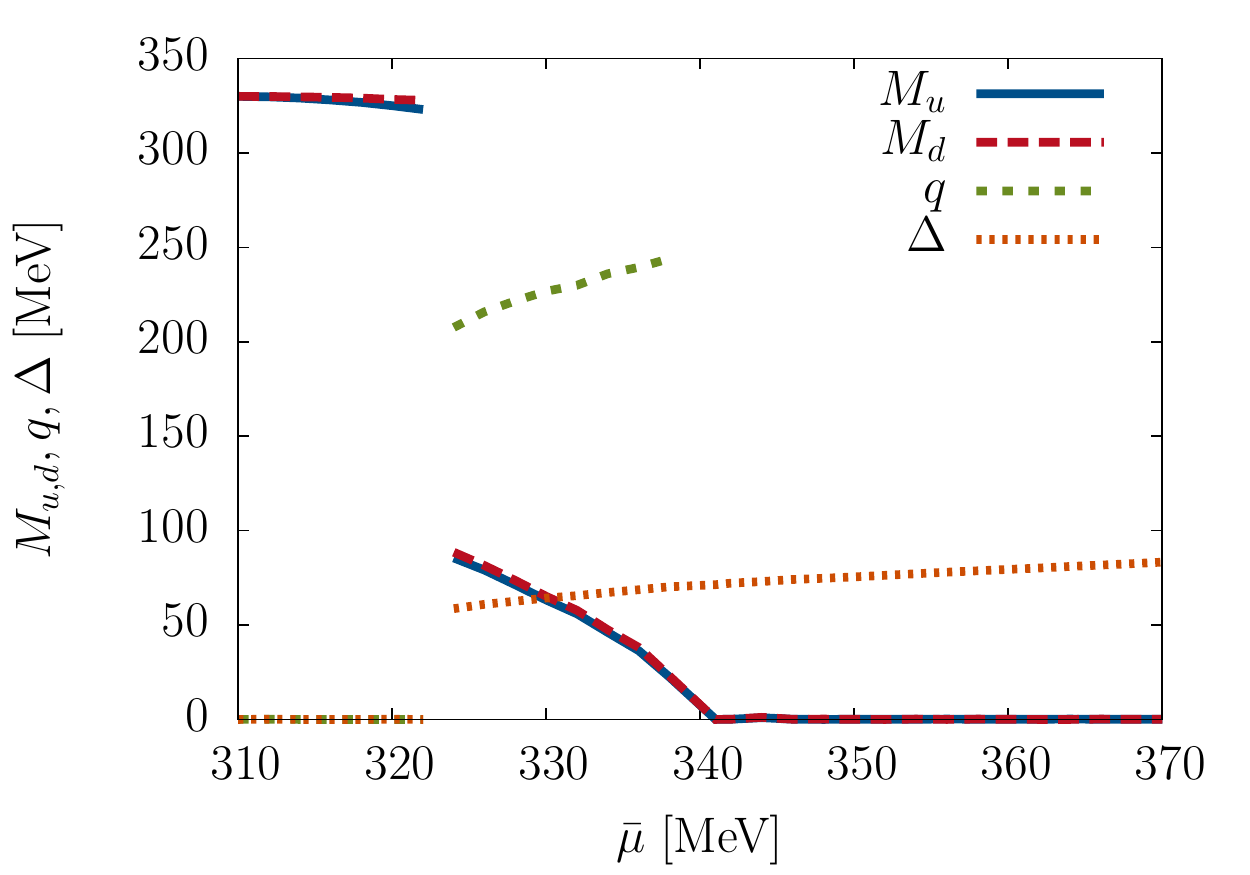}
  \includegraphics[width=0.48\textwidth]{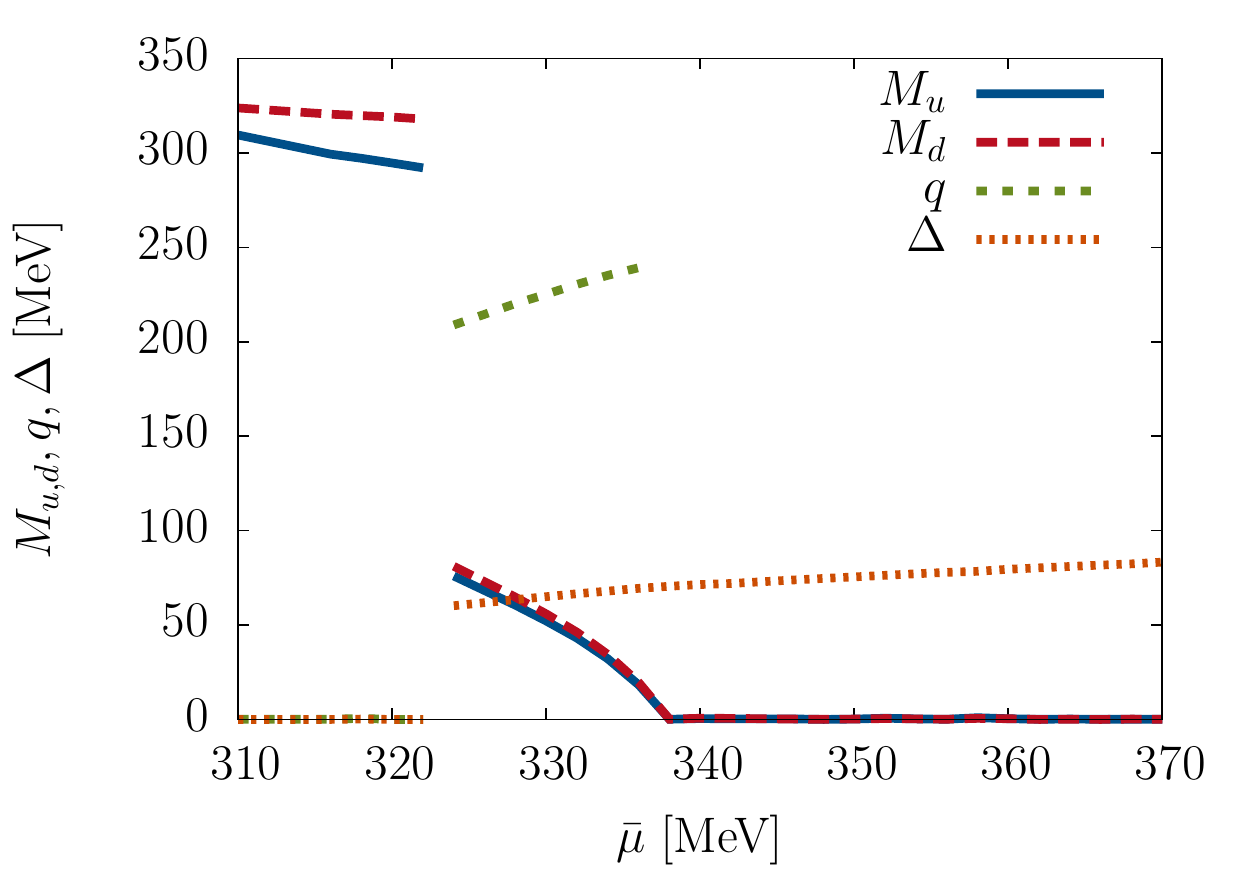}
\caption{Energetically favored values of the variational parameters at $\mu_I=40$ (left) and 80 MeV (right) for $H=G/2$ and $T=0$, as a function of $\bar\mu$. \label{fig:orderparams4080}}
\end{figure}

After having discussed the $\mu_I=0$ case, we now move to the case of asymmetric quark matter. 
It is known that a nonzero isospin chemical potential should disfavor BCS pairing, eventually leading to the breakup of the diquark condensate once it surpasses a critical value. While we expect this value to be close to the Chandrasekhar-Clogston (CC) limit,
 the two might differ a bit due to the strong coupling dynamics and the interplay with chiral symmetry breaking.
In \Fig{fig:orderparams4080} we show the behavior of the order parameters as a function of $\bar\mu$ for a diquark coupling $H = G/2$ and $\mu_I =40$ and 80 MeV. We observe that for these moderate isospin asymmetries the inhomogeneous phase slightly shrinks as $\mu_I$ grows, although the effect is very small. We also note that, while a larger 
isospin asymmetry affects the splitting between the amplitudes for the mass modulations associated with the two different flavors, the values of $q$ and $\Delta$ remain almost unmodified by the change in $\mu_I$, an aspect which will be investigated in more detail at the end of this section. 

\begin{figure}
 \centering
  \includegraphics[width=0.48\textwidth]{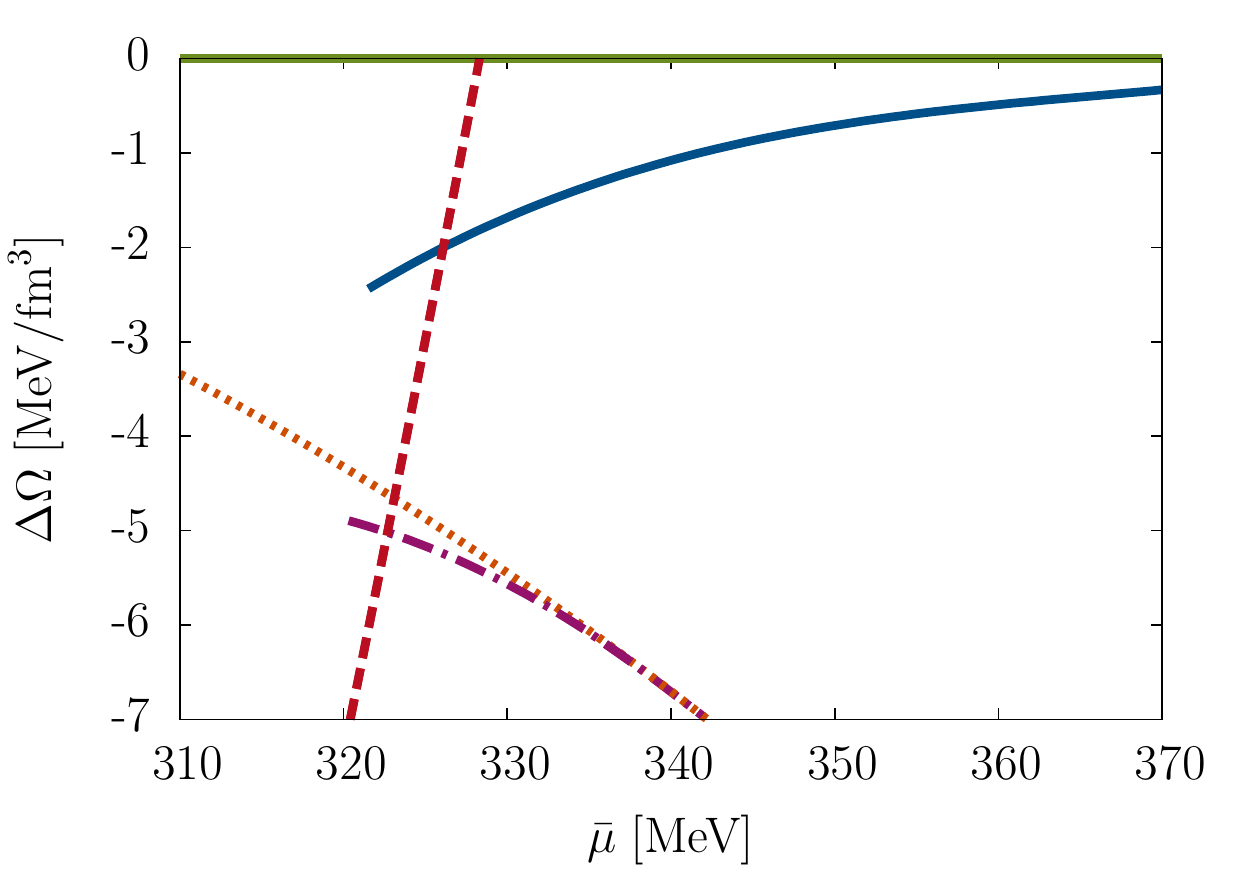}
 \includegraphics[width=0.48\textwidth]{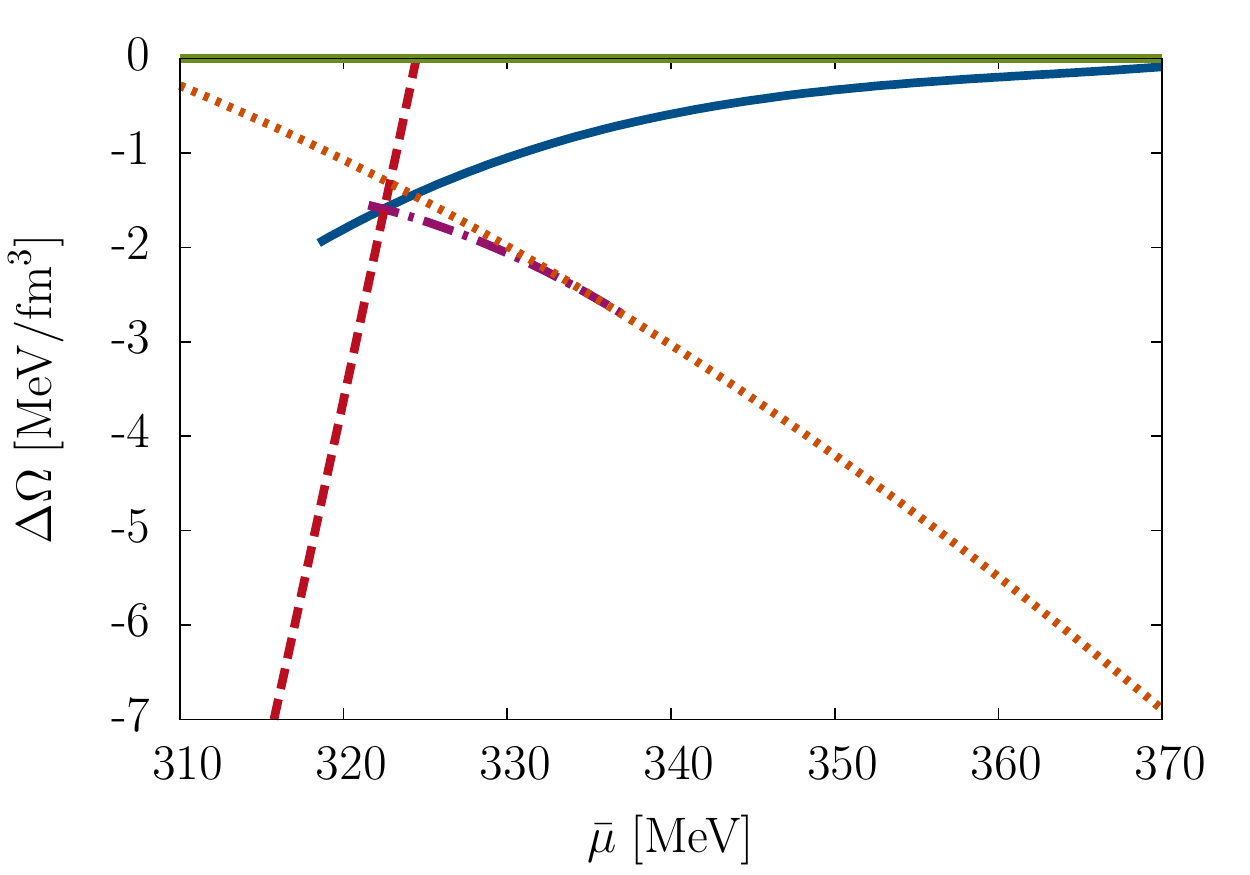}
\caption{Thermodynamic potential (normalized to the restored phase) as a function of the average quark chemical potential for $T=0$, $H=G/2$, $\mu_I=40$ (left) and 80 MeV (right). Dashed red line: homogeneous chiral condensates, solid blue line: inhomogeneous chiral condensates, dotted orange line: only diquark condensation,
dash-dotted purple: coexistence phase with  simultaneous inhomogeneous chiral symmetry breaking and homogeneous color-superconductivity. 
 \label{fig:omegas4080}}
\end{figure}

\begin{figure}
 \centering
  \includegraphics[width=0.48\textwidth]{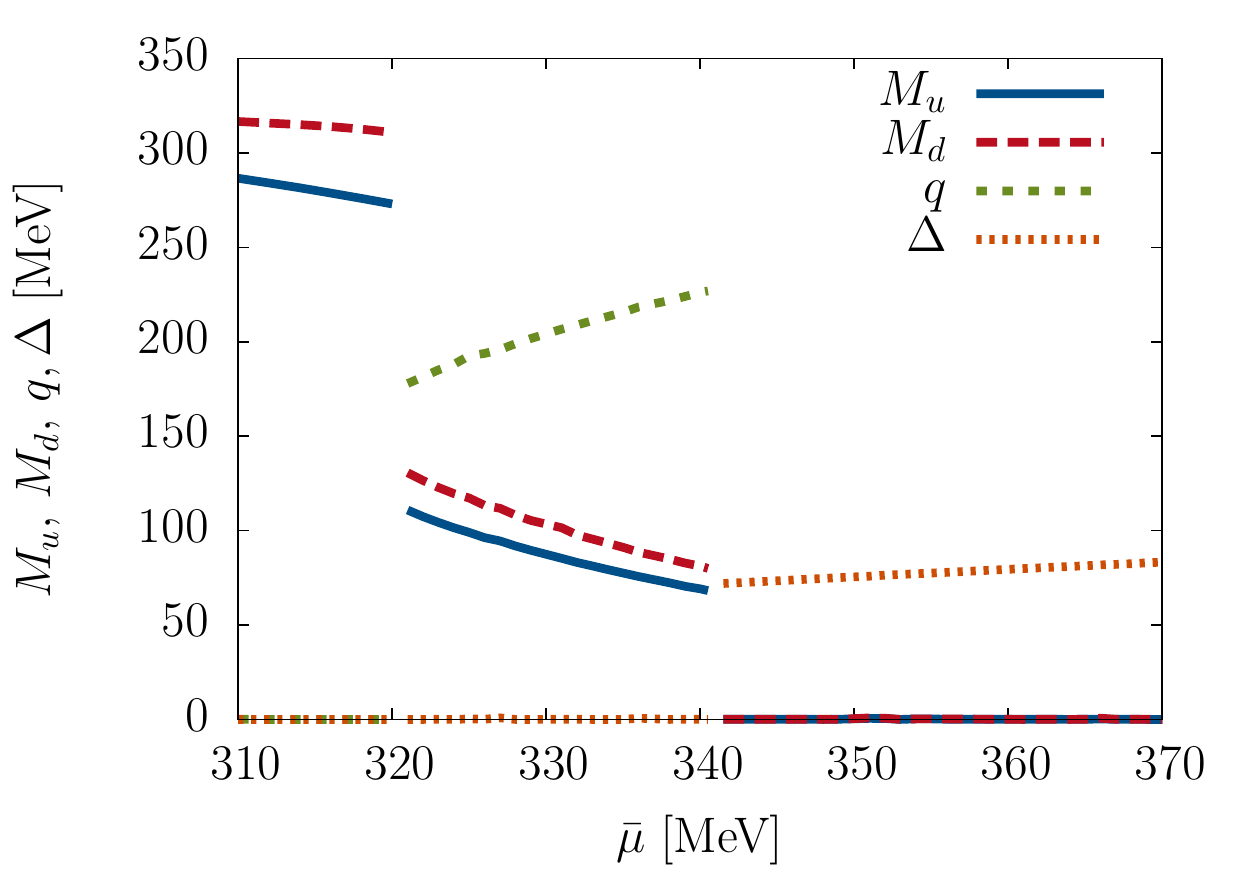}
  \includegraphics[width=0.48\textwidth]{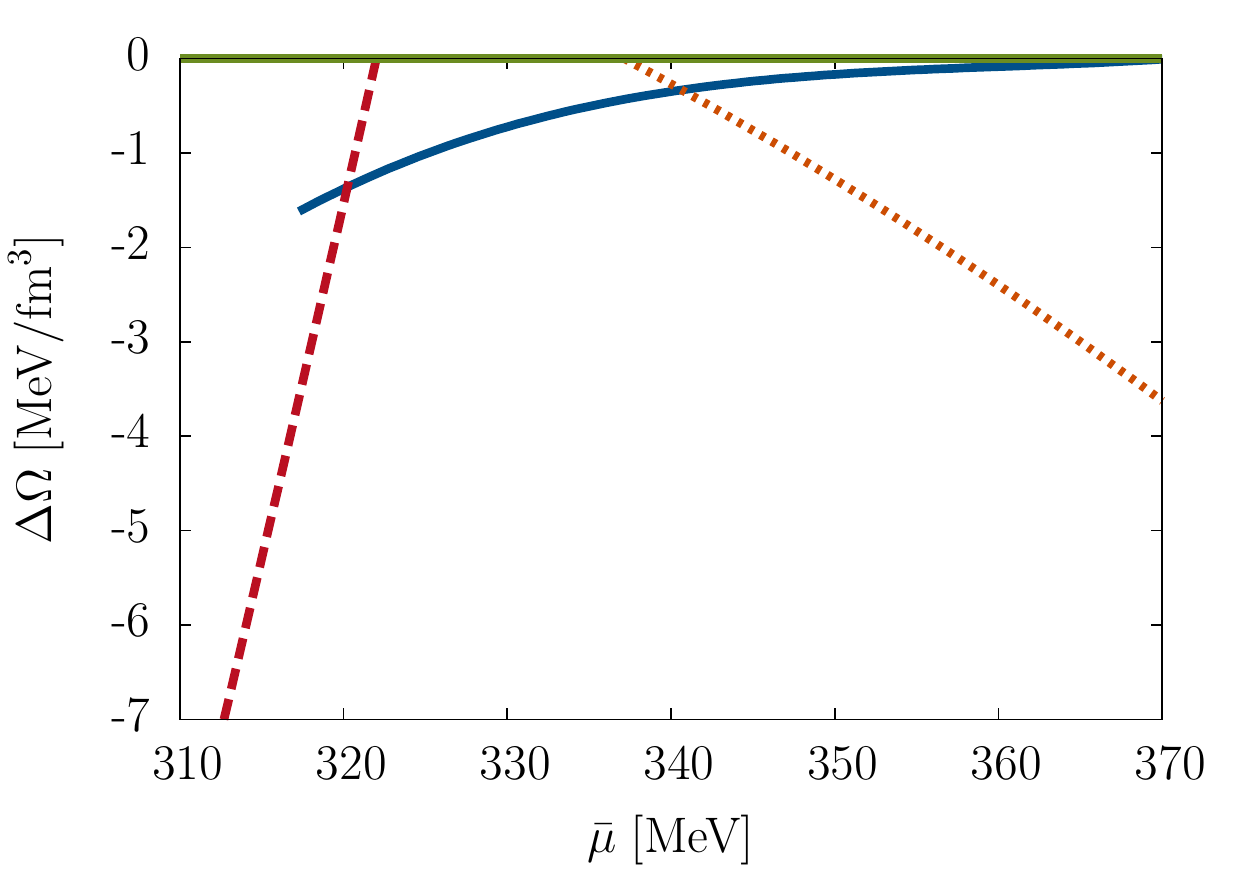}
\caption{Left: Energetically favored values of the variational parameters 
 as a function of the average quark chemical potential for $T=0$, $H=G/2$, $\mu_I=100\;\text{MeV}$.
 Right: Thermodynamic potential normalized to the restored phase. Dashed red line: homogeneous chiral condensates, solid blue line: inhomogeneous chiral condensates, dotted orange line: only diquark condensation. The coexistence phase line lies above the one associated to the chirally restored solution and is not shown here.
  \label{fig:orderparams100}}
\end{figure}

  The effect of $\mu_I$ is perhaps more visible in the behavior of the free energies for the different phases, which are shown in \Fig{fig:omegas4080}. There we observe how the isospin chemical potential shifts the 2SC curves upwards, making these solutions less favored. In particular, for $\mu_I = 80$ MeV we approach the point where the solutions with nonzero diquark gaps are almost degenerate with the $\Delta=0$ ones.
This in turn suggests that significant deviations might start to occur as the isospin chemical potential further increases.
Indeed, this is the case and for $\mu_I$ slightly above this value we observe that as $\bar\mu$ increases the system undergoes a first phase transition from the homogeneous chirally broken phase to an inhomogeneous phase where $M_f \neq 0$, $q \neq 0$ and $\Delta = 0$, followed by another first-order transition into the coexistence phase, and finally reaches the pure 2SC phase. As $\mu_I$ further increases, the pure 2SC phase gets pushed to even higher $\bar\mu$ and there is no coexistence region.
We see this in \Fig{fig:orderparams100}, where the order parameters and the free energy associated to the different solutions for $\mu_I =100$ MeV are plotted: the coexistence phase 
disappears and the 2SC condensate appears only after a first-order transition at which the chiral condensate sharply drops to zero.
Eventually, if the isospin asymmetry is raised further, the chiral condensate melts completely before the onset of the 2SC phase, so that a chirally restored window with no diquark gap separates the two.

\begin{figure}
 \centering
 \includegraphics[width=0.48\textwidth]{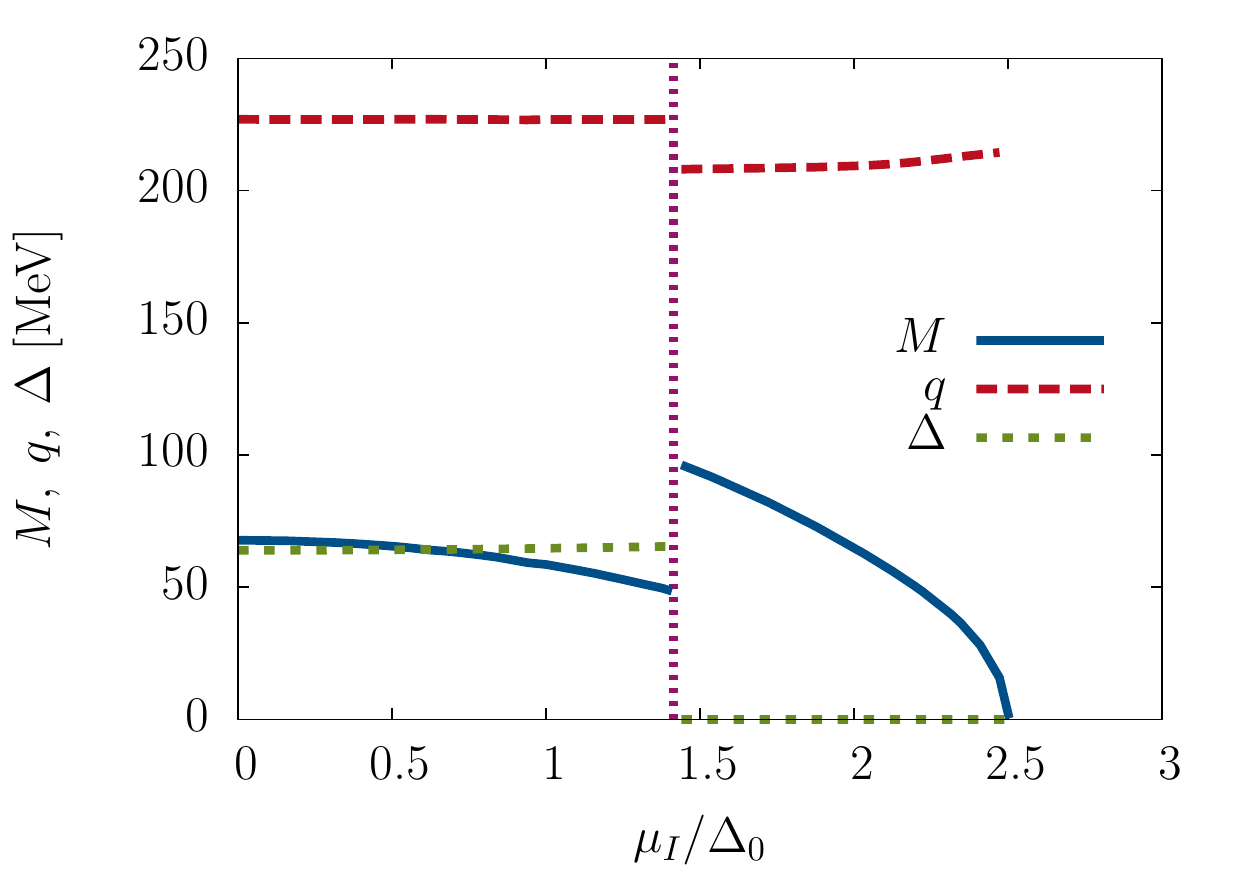}
 \includegraphics[width=0.48\textwidth]{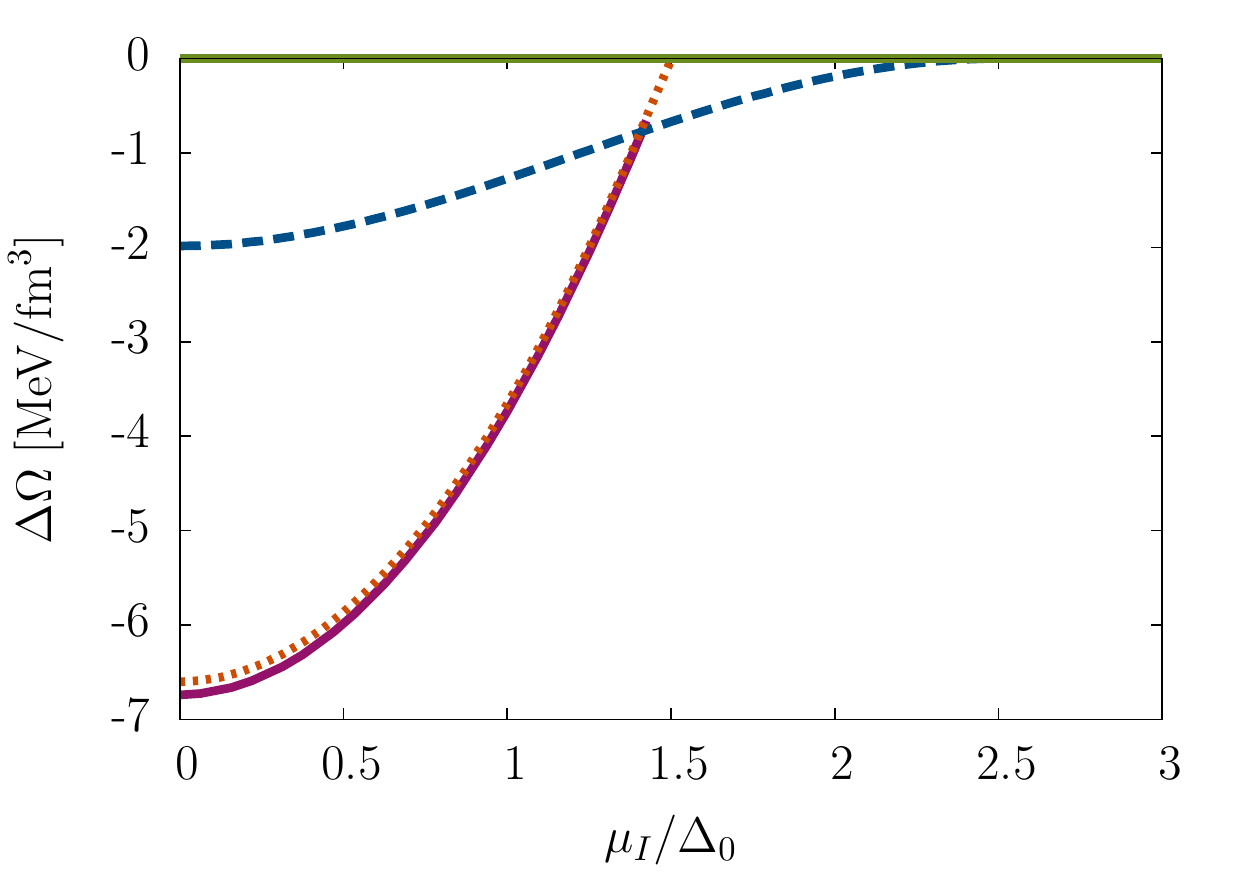}
\caption{Left: Variational parameters at $\bar\mu = 330$ MeV and $H=G/2$, $T=0$, as a function of $\mu_I$. The vertical line indicates the Chandrasekhar-Clogston limit. 
Right: Thermodynamic potential corresponding to the different solutions. Solid purple line: coexistence phase with simultaneous occurrence of inhomogeneous chiral symmetry breaking and homogeneous color-superconductivity, dashed blue line: inhomogeneous chiral condensates, dotted orange line: pure 2SC phase.\label{fig:orderparamsmuI}}
\end{figure}

Complementary to these results, in \Fig{fig:orderparamsmuI} we show the behavior of the order parameters and the free energies for a fixed average chemical potential $\bar\mu = 330 $ MeV, as function of $\mu_I$. For sake of simplicity here we consider a more restrictive CDW ansatz which can be obtained by \Eq{eq:Mofx} by imposing $M_u = M_d = M$ (we expect this restriction to be not too severe, particularly at lower $\mu_I$) . 
We observe that in the coexistence phase the diquark gap shows a very weak dependence on $\mu_I$, changing by about 2 MeV from $\mu_I =0 $ to the CC limit $\mu_I = \sqrt{2} \Delta_{0} \approx 90$ MeV (where we define $\Delta_{0}$ as the value of the diquark gap for $\mu_I = 0$). 
This dependence should not be present in a pure 2SC phase \cite{Bedaque:1999nu}, and we interpret it as an effect of the interplay with inhomogeneous chiral symmetry breaking\footnote{Indeed, for a pure 2SC solution in our model the gap remains constant with $\mu_I$.}. We observe that the value of the wave number $q$ associated to the chiral condensate also remains practically constant, while the amplitude of the chiral condensate on the other hand gradually melts. 
Close to the CC value there is a first-order transition 
after which BCS condensation becomes disfavored, leading to a phase where chiral symmetry is inhomogeneously broken and $\Delta = 0$. At this transition, the values of $M$ and $q$ jump, and then gradually melt to zero. Within this parameter set the chiral condensate melts completely at $\mu_I \approx 160$ MeV, a value which at any rate lies beyond the realm of validity of this model since charged pion condensation is expected to set in before then. 

\begin{figure}
 \centering
\includegraphics[width=0.48\textwidth]{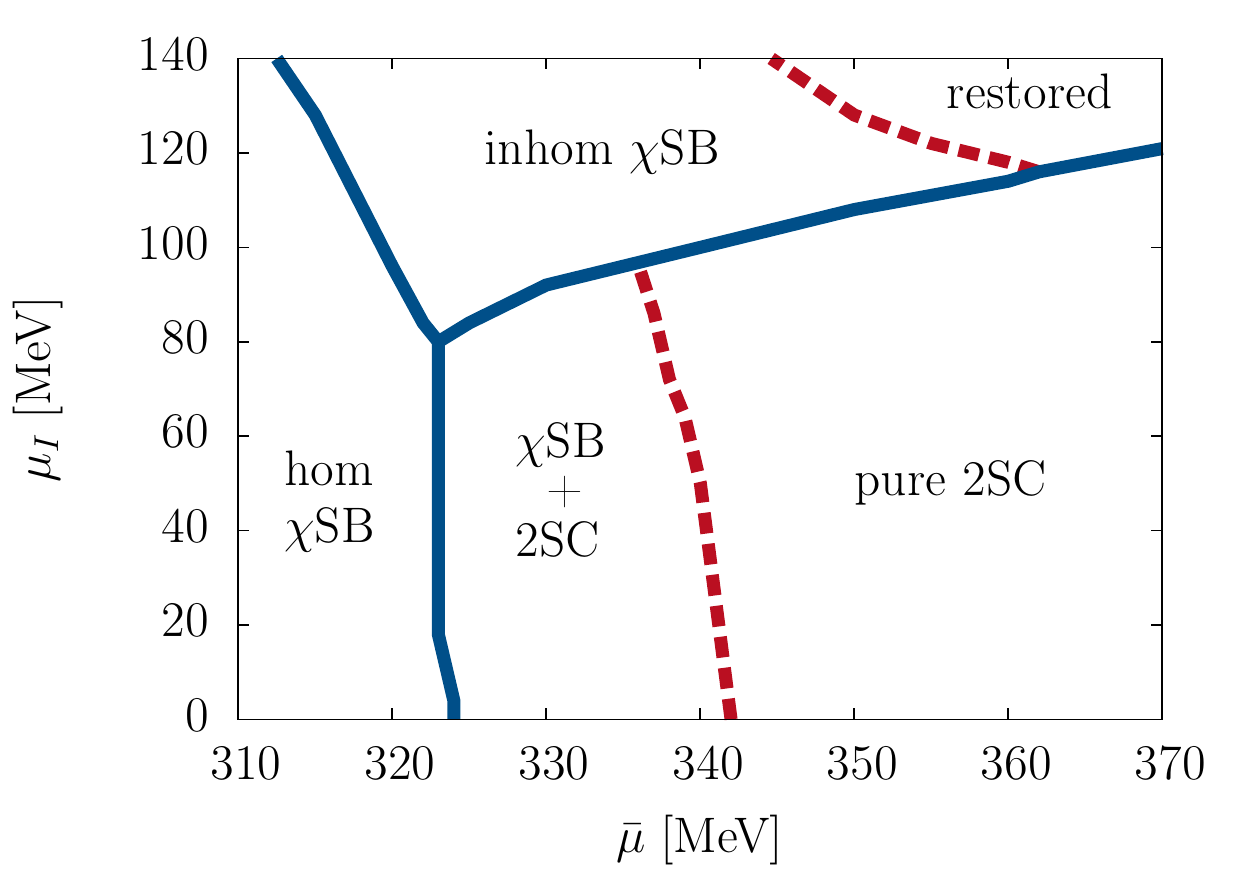}
\caption{Phase diagram in the  $\bar\mu$-$\mu_I$--plane at $T=0$ and $H=G/2$. Solid blue lines denote first-order phase transitions, while dashed red lines indicate second-order ones.  The label ``hom $\chi$SB'' characterizes a region where $M\neq0,\;q=\Delta=0$, ``$\chi$SB+2SC'' denotes a region with $M\neq0,\;q\neq0,\;\Delta\neq0$, while ``inhom $\chi$SB'' corresponds to a region characterized by $M\neq0,\;q\neq0,\Delta=0$ and ``pure 2SC'' labels a domain where $M=0,\;\Delta\neq0$.
\label{fig:pdmumui}}
\end{figure}

Finally, in \Fig{fig:pdmumui} we show the $\bar\mu - \mu_I$ phase diagram for $H = G/2$, showing the extension of the coexistence phase from low to moderate isospin chemical potentials. At high $\bar\mu$ the pure 2SC phase becomes thermodynamically favored, as expected.  For high $\mu_I$ instead the diquark condensation becomes disfavored, so that for a given range in $\bar\mu$ at high isospin chemical potentials the coexistence phase ends and only inhomogeneous chiral symmetry breaking remains.

\section{Summary and outlook}

We investigated isospin-asymmetric two-flavor quark matter within an extended NJL model, focusing in particular on the interplay between inhomogeneous chiral symmetry breaking and color-superconductivity.  For simplicity, we assumed the diquark gap to be spatially homogeneous, a constraint which becomes increasingly restrictive as higher isospin asymmetries are considered. We found that at vanishing isospin chemical potential, as the density of the system increases there is a first-order phase transition from the normal homogeneous chirally broken phase to a coexistence region where inhomogeneous chiral symmetry breaking and a nonzero homogeneous 2SC gap are simultaneously present. 
The presence of a diquark gap leads to a gradual melting of the chiral condensate, so that beyond a critical baryon chemical potential there is a second phase transition to a pure 2SC phase. As a consequence, the inhomogeneous ``continent'' disappears. 
The size of the coexistence region depends on the strength of the diquark coupling, and for larger values such as $H=3/4 G$ the pure 2SC phase becomes thermodynamically favored
over all inhomogeneous solutions. In order to determine whether such a window can still exist for larger couplings, it would be interesting to consider inhomogeneous solutions which are considered to be thermodynamically more favored than simple plane waves \cite{Nickel:2009wj,Carignano:2012sx}, although this would significantly complicate calculations.

We found that for small isospin asymmetries, the size of this coexistence window is almost unaffected by the value of the isospin chemical potential, while for larger values of $\mu_I$ the diquark condensation gets pushed to higher chemical potential, allowing for the existence of an intermediate phase where chiral symmetry is inhomogeneously broken but no color-superconducting gap is formed.  Of course this scenario should be taken with care, since in this work we neglected the possibility of formation of crystalline color-superconducting condensates, which are expected to appear in imbalanced systems for sufficiently high asymmetries. Is it then very likely that the real ground state of dense matter under these conditions is a  coexistence phase where both the chiral condensate and the diquark gap are spatially modulated.  As a matter of fact, if the chiral condensate becomes inhomogeneous and induces a spatial dependence on the density of the system, it is entirely possible that color-superconducting islands start forming in the high density regions of the system already at zero isospin chemical potential. This is a fascinating possibility that would require however a numerically intensive computation, e.g. by combining the numerical method developed in  \cite{Nickel:2008ng} for inhomogeneous color-superconductivity with the one described in \cite{Carignano:2012sx} for crystalline chiral condensates.

Ultimately, our aim is to provide a realistic description of matter in the core of compact stars. For this, our model should be extended to incorporate the effects of strong magnetic fields and vector interactions, both of which are expected to have strong effects on the size of the inhomogeneous phase and its properties, see \cite{Frolov:2010wn,Nishiyama:2015fba} and \cite{Carignano:2010ac}, respectively. Whether these effects extend the coexistence region of chiral symmetry breaking and color-superconductivity or simply push the diquark onset to higher chemical potentials is a very interesting question which should be addressed within explicit model calculations.

\section*{Acknowledgments} 
We thank the organizers for the very stimulating atmosphere at the workshop as well as for financial support. 
We are also grateful to M. Buballa and M. Schramm for useful discussions.

\bibliographystyle{JHEP}
\bibliography{biblio}


\end{document}